\documentclass[aps,pre,reprint,groupedaddress,showpacs]{revtex4-1}

\usepackage{amsmath}
\usepackage{graphicx}
\begin{document}

\title{Density-Independent Model of Self-Propelled Particles}

\author{Daniel Schubring}

\author{Paul R. Ohmann}
\affiliation{University of Saint Thomas\\
2115 Summit Avenue, Saint Paul, MN 55105, USA}

\begin{abstract}
We examine a density-independent modification of the Vicsek model in which a particle interacts with neighbors defined by Delaunay triangulation.  To feasibly simulate the model, an algorithm for repairing the triangulation over time was developed. This algorithm may also be applied to any time varying two-dimensional Delaunay triangulation. This model exhibits a continuous phase transition with noise, and a distinct set of critical exponents were measured which satisfy a hyperscaling relationship. The critical exponents are found to vary between a low and high velocity regime, but they are robust under the inclusion of a repulsive interaction. We present evidence that the correlation length approximately scales with the size of the system even in the ordered phase. 
\end{abstract}


\maketitle

\section{Introduction}
The study of collective motion is a rapidly growing field which employs techniques from physics to examine systems such as crowds, traffic, and flocks of animals \cite{review}. These systems all lead to the emergence of group behavior which is not easily predictable from the behavior of the individuals.

One of the hallmarks of complex systems, such as those studied in this field, is the notion of universality--- certain quantities may be calculated which are quantitatively the same across many systems. If a complicated but realistic model is in the same universality class as a simple model, some features of the realistic model may be calculated using the simpler model.

Within the last five years a landmark observational study of starling flocks has been published, suggesting many new ideas for models of collective motion \cite{topdistance,starling,correlations}. There has also been a recent model inspired by this research that has replicated some important features of the observed behavior \cite{starmodel}. 

However, in contrast to the above model we wish to examine a model which is as simple as possible while still capturing the important qualitative features observed in the starling research. Following an earlier cohesive model \cite{gregoire03}, we will examine a modified Vicsek model in which `topological' nearest neighbor interactions are based on a Voronoi tesselation.

More recently, a similar Voronoi neighbor model was also studied by Ginelli and Chat\'{e} \cite{ginelli}, and hydrodynamic equations for self-propelled particles with a nearest neighbor interaction were also developed \cite{hydrodynamic}. In agreement with the model of Ginelli and Chat\'{e}, we will find evidence of a continuous phase transition and calculate a complete set of critical exponents. The critical exponents are found to vary with a velocity parameter, though they appear to be stable in both a low and high velocity regime. 

We also present tentative evidence that the correlation length scales with the size of the system, even in the ordered phase. Such behavior could explain the observation of the scaling of correlation length in starling flocks \cite{correlations}, which of course involve much smaller system sizes and sampling sizes than the computer simulations investigated in this paper. Finally, we present the algorithm developed to update the Voronoi tesselation over time, which can significantly reduce the computational cost of simulating models of this type --- and which may be of interest to researchers in unrelated fields.

\section{The Vicsek Model}
We will begin with a discussion of the Vicsek model \cite{vicsek} --- a model which despite its simplicitly exhibits much interesting behavior. In this model, $ N $ individuals move with the same fixed speed $v$ and adjust their direction to match the average velocity of their neighbors. The neighborhoods have a fixed length scale $ \Delta L $\@. The neighborhoods may simply be defined as all individuals within a radius of $ \Delta L $, but implementations with neighborhoods based on a fixed lattice lead to the same critical behavior \cite{vicsek2}. The individuals interact in a two dimensional square of linear size $ L \Delta L $ with periodic boundary conditions. The interaction rule takes place at discrete time steps of length $ \Delta t $. In the following, we will fix the units by setting both $ \Delta L $ and $ \Delta t $ equal to one. At each time step, the new velocities $ v_{i}(t + 1) $ are first calculated for all individuals, and then the position of the ith individual is updated as
\begin{equation}
x_{i}(t + 1) = x_{i}(t) + v_{i}(t).
\label{posupdate}
\end{equation} 

Another possible version of this update rule uses $ v_{i}(t + 1) $ in Eq.\eqref{posupdate} rather than $ v_{i}(t) $. The two cases are known as forward and backward update, respectively. The difference between the two update rules may seem trivial, but the forward update rule may facilitate the appearance of a phase of traveling waves \cite{gregoire08}. There has been much debate on whether these traveling bands, which lead to a discontinuous phase transtion are an essential aspect of the Vicsek model or an artifact based on boundary conditions \cite{aldana}. Since the backward update rule does not appear to lead to this phase \cite{albano09} we will stick to this rule throughout.

The direction of the velocity of the ith individual $ \theta _{i}(t + 1) $ is updated to match the argument of the average velocity in its neighborhood $ N_{i}(t) $ in the presence of a random perturbation. 
\begin{equation}
\theta _{i}(t + 1) = \langle\theta(t)\rangle_{N_{i}} + \eta \xi _{i}(t)
\label{velupdate}
\end{equation} 
Here $ \xi _{i}(t) $ is a uniformly distributed random variable on the interval $ [0, 1] $, and $ \eta $ is a global parameter of the model. Note that this is a so-called `angular' noise interaction rather than a `vectorial' noise interaction, which again has been shown to lead to a discontinuous phase transition \cite{gregoire08}. In particular note that the model of Ginelli and Chat\'{e} \cite{ginelli} uses a vectorial noise interaction with a forward update rule, in contrast to the model in this paper.

In order to study the phase transition as $ \eta $ is varied between $ 2\pi $ and $ 0 $, we introduce the instantaneous order parameter $ \phi(t) $.

\begin{equation}
\phi(t) = \frac{1}{N v} \big\| \sum_i v_i \big\|
\label{phi}
\end{equation} 

We will be interested in the ensemble average of $ \phi $ as a function of the global parameters, which we will denote with the variant symbol $ \varphi(\eta) $. To calculate $ \varphi(\eta) $ we will adopt the ergodic hypothesis that the average over time converges to the ensemble average, $ \varphi=\lim_{T\rightarrow\infty}\frac{1}{T} \sum_{t=0}^{T} \phi(t) $.

Numerous papers have affirmed that the Vicsek model demonstrates a continuous phase transition as $ \eta $ is varied \cite{aldana,albano08,albano09} (at least if the conditions are such that the traveling band effect does not appear). In the thermodynamic limit, $ \varphi $ is zero above the critical point $ \eta_C $, and near the critical point follows the scaling law
\begin{equation}
\varphi \sim [\eta_{C} - \eta]^\beta.
\label{beta}
\end{equation}
Here, $ \beta $ is the order parameter critical exponent. We will also be interested in the critical exponent $ \nu $, which describes the divergence of the correlation length at $ \eta_C $, and the susceptibility exponent $ \gamma $. The latter describes the divergence of the susceptibility $ \chi $, which in this context we will define in terms of the variance $ \sigma^2 $ of $ \phi $.
\begin{equation}
\chi (\eta,N) = N\sigma^2(\eta,N)
\label{chi}
\end{equation}
For feasible values of $ N $, the scaling laws described by these exponents ---and even the location of the critical point itself--- are obscured by strong finite size effects. The problem of measuring these exponents can be dealt with by finite-size scaling methods, as demonstrated by Baglietto and Albano \cite{albano08}. For equilibrium systems, the critical exponents must also satisfy the hyperscaling relationship
\begin{equation}
d\, \nu - 2 \beta = \gamma,
\label{hyperscaling}
\end{equation}
where $ d = 2 $ is the dimension of the space. The exponents calculated by Baglietto and Albano also satisfy this relationship, giving credence to the use of these methods for the non-equilibrium Vicsek model.
\section{A Density-Independent Model}
In the standard Vicsek model, density is also an important parameter. The position of the critical point has been found to depend on the global density $ \rho = N / L^2 $ according to $ \eta_C \sim \sqrt{\rho} $ \cite{vicsek2,albano08}. Furthermore, the local density of individuals is correlated with the local order \cite{gregoire08}. This can be understood simply through the fact that the number of neighbors involved in the update rule increases as the density increases. 

However, there is reason to suspect that density may not always be an important parameter for modeling collective motion in nature. A large scale empirical study of starling flocks provides a number of interesting observations from which to form a new model of collective motion \cite{topdistance,starling,correlations}. In particular, the starling flocks were observed to have a wide variety of densities and nearest neighbor distances while having similar behavior in other respects \cite{starling}. Further, it was argued that the starlings always interact with a certain number of closest neighbors regardless of the metric distance between them--- a property termed \emph{topological distance} \cite{topdistance}. Aiming to build a simple model based on topological distance, we sought to replace the fixed length neighborhoods of the Vicsek model with something invariant under a change of length scale.

One possible option might be to redefine the neighborhood $ N_i $ as the $ n_c $ closest neighbors to the individual $ i $, where $ n_c $ is a global parameter. The authors of the starling research suggested such an approach and even measured $ n_c  = 6.5 \pm 0.9 $ \cite{topdistance}. We ran a simulation based on this idea which suggested that the individuals would form dense clusters which would only weakly interact with each other through collisions. This behavior can be understood since the individuals already have $ n_c $ neighbors within their cluster which are closer than the individuals in any other cluster except during rare collisions. To overcome this `geometric instability' a recent cohesive model of this type has required a minimum angular resolution between the $ n_c $ neighbors \cite{sbalanced}.

Following an earlier precedent \cite{gregoire03}, we instead looked to define the nearest neighbors in terms of the Voronoi tessellation \cite{voronoi}. Here each individual is associated with a cell of points which are closer to that individual than any other. An individual’s neighbors are defined as those individuals in adjacent cells.

Using these Voronoi neighborhoods, particles on the boundary of a cluster may still interact with particles outside the cluster even if they are at a greater metric distance. Indeed, if we maintain periodic boundary conditions, individuals on one side of a large flock may have individuals on the opposite side of the flock as neighbors. So we would expect the large scale behavior of this model to be quite different from either the Vicsek model or the model based on the $ n_c $ parameter. In particular, we would expect that persistent locally dense `flocks' will not form in the absence of an explicit cohesion term in the interaction. Indeed, in a similar model developed independently to that of this paper, Ginelli and Chat\'{e} have previously observed this qualitative difference from the metric Vicsek model \cite{ginelli}.

When we define the neighborhoods in terms of topological distance, we lose the length scale $ \Delta L $ with which we defined our units. In the following, we instead choose to set our units by requiring that $ \rho = 1 $. To be clear, if we have $ N $ individuals, we set the length of the two-dimensional system space as $ \sqrt{N} $ and express the velocity $ v $ in terms of this scale.

\section{ Delaunay Triangulation}
\label{delaunay}

A number of standard algorithms exist for computing the Voronoi tessellation of a bounded set of points \cite{voronoi}. However, there are two additional considerations in implementing the new model which are not as commonly dealt with. First of all, since the system has no natural boundary, we must be able to extend the Voronoi tessellation over periodic boundary conditions. The second point we will consider is that in the low-velocity regime, we do not expect an individual’s neighbors to change significantly over successive time steps. In this regime, it may be much more efficient to use an algorithm that updates the existing neighborhoods over time to maintain the Voronoi property.

\begin{figure}[!t]
\includegraphics[width=.4\textwidth]{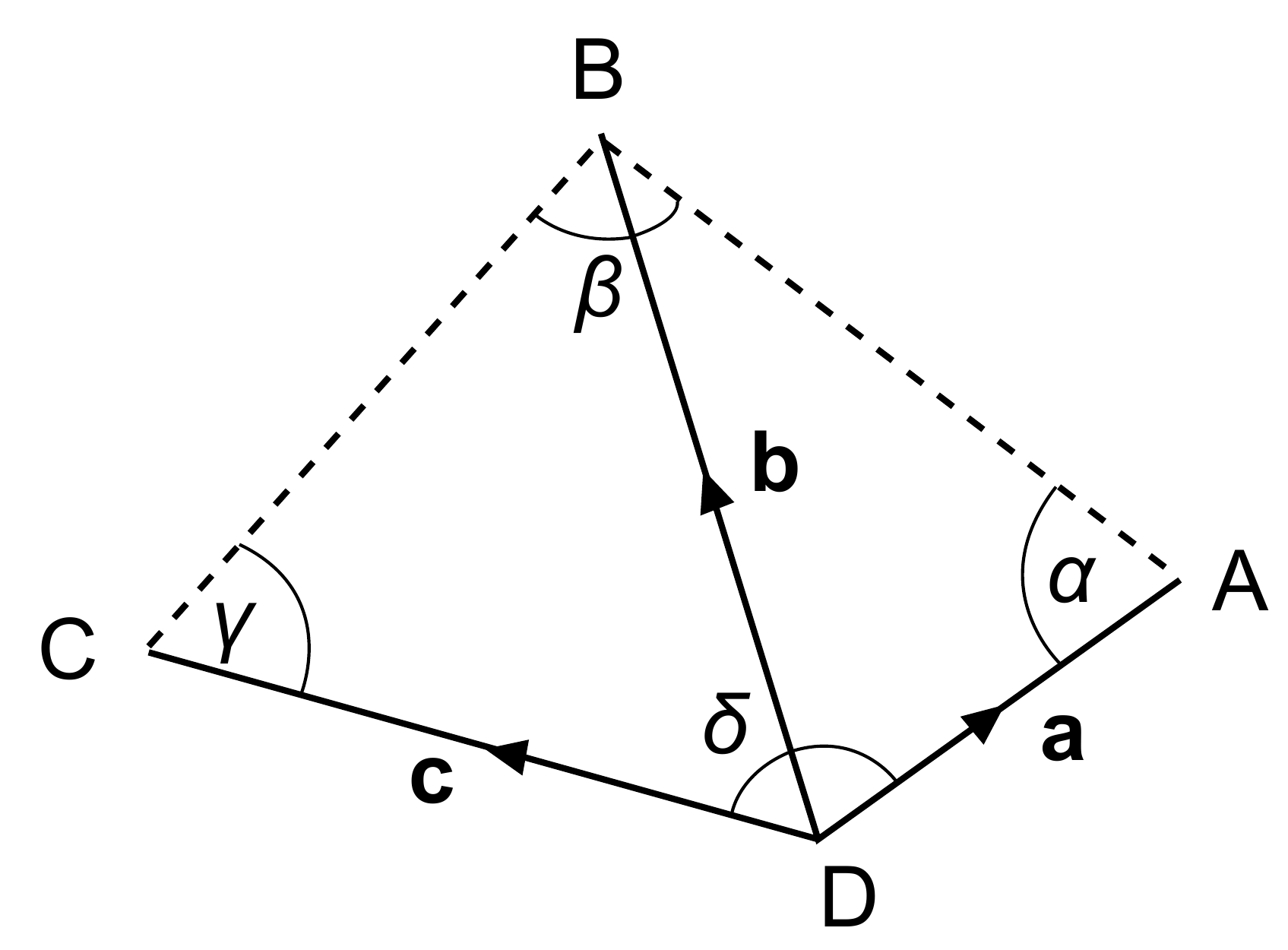}
\caption{The geometry of the edge $ \overline{BD} $ and its neighboring triangles in the Delaunay triangulation.
\label{figtriangle}}

\end{figure}

In implementing the model, it is helpful to deal with the dual of the Voronoi tessellation--- the Delaunay triangulation. Here instead of considering the cells themselves, we consider the graph consisting of the edges connecting neighbors. A general two-dimensional triangulation is a Delaunay triangulation if and only if each edge satisfies a certain condition considered below.

Consider the edge $ \overline{BD} $, along with its two neighboring triangles in Fig. \ref{figtriangle}. $ \overline{BD} $ has the Delaunay property if its opposite angles satisfy the inequality $ \alpha + \gamma \leq \pi $. Note that if $ \overline{BD} $ fails this inequality, the \emph{flipped} edge $ \overline{AC} $ will satisfy it. This suggests a simple algorithm for converting a general triangulation to a Delaunay triangulation: test whether each edge satisfies this inequality, and replace the edge with the flipped edge if it fails. This algorithm does indeed always terminate in the Delaunay triangulation \cite{voronoi}. In the implementation of this algorithm, it is more efficient to calculate the inequality equivalently in terms of the vectors $ \mathbf{a} $, $ \mathbf{b} $, and $ \mathbf{c} $
\begin{equation}
\|\mathbf{a}\|^2 \|\mathbf{b}\times\mathbf{c}\|
	+\|\mathbf{b}\|^2 \|\mathbf{c}\times\mathbf{a}\|
		+\|\mathbf{c}\|^2 \|\mathbf{a}\times\mathbf{b}\| \geq 0
\end{equation}
By using this flipping algorithm, we can simplify the task of finding the full Delaunay triangulation to that of finding any valid triangulation over periodic boundary conditions. We first construct any bounded triangulation of the individuals whose boundary is also the convex hull. Any of the standard algorithms to construct the Delaunay triangulation are suitable here, but since the triangulation need not be a valid Delaunay triangulation at this point, we use the S-Hull algorithm \cite{shull}.

To extend this triangulation over the periodic boundary conditions we next connect the points on the convex hull to each other. We first connect the points with maximum and minimum y-values to themselves, creating degenerate edges which will require special handling later in the flipping algorithm. Then we stitch the left and right sides of the convex hull together, the details of which are in Appendix \ref{boundary}.

\begin{figure}[!b]
\includegraphics[width=.4\textwidth]{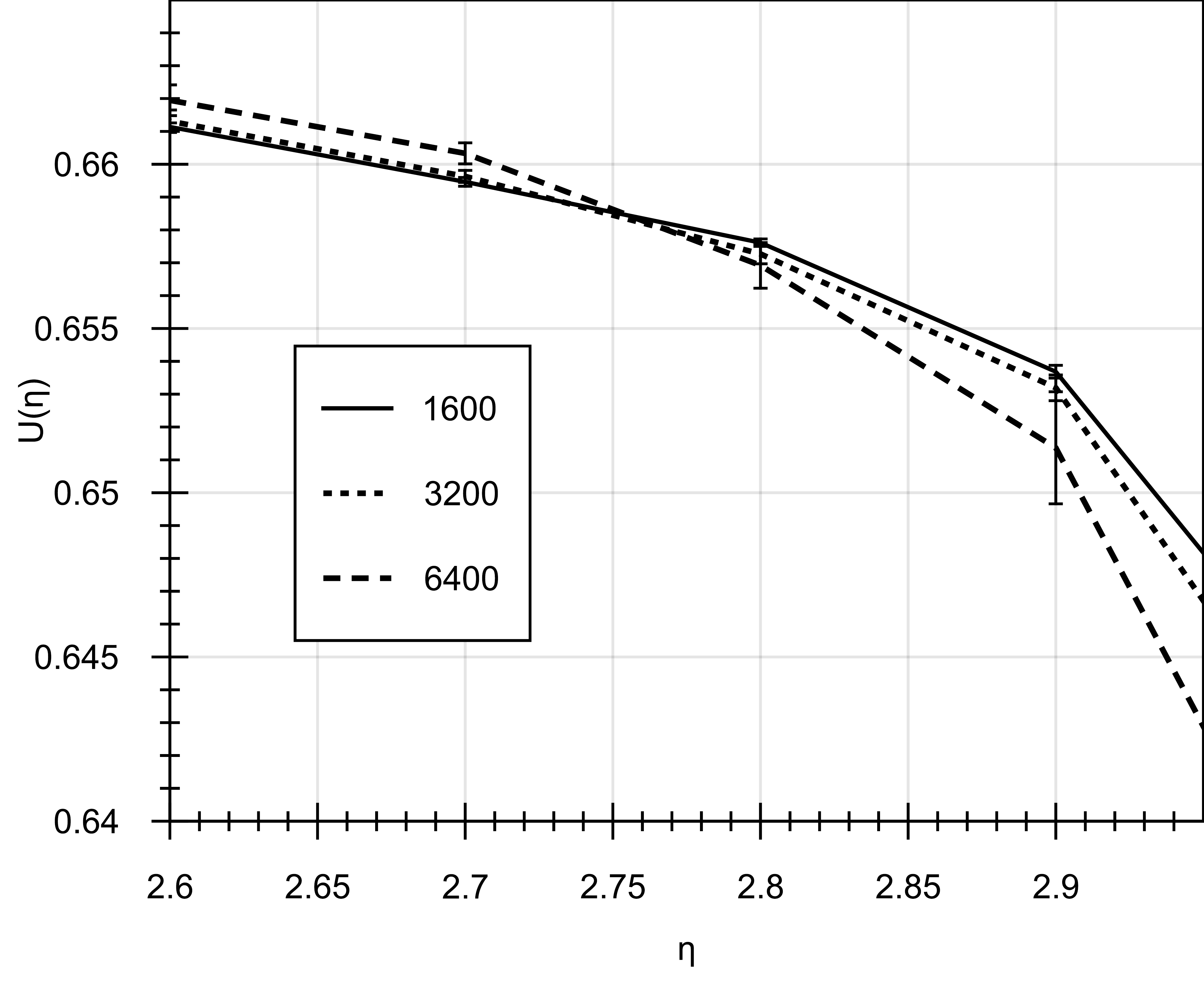}
\caption{The Binder cumulants of three different system sizes appear to cross near $\eta_C = 2.75$ .
\label{figbinder}}
\end{figure}

\begin{figure*}[!t]
\includegraphics[width=.45\textwidth]{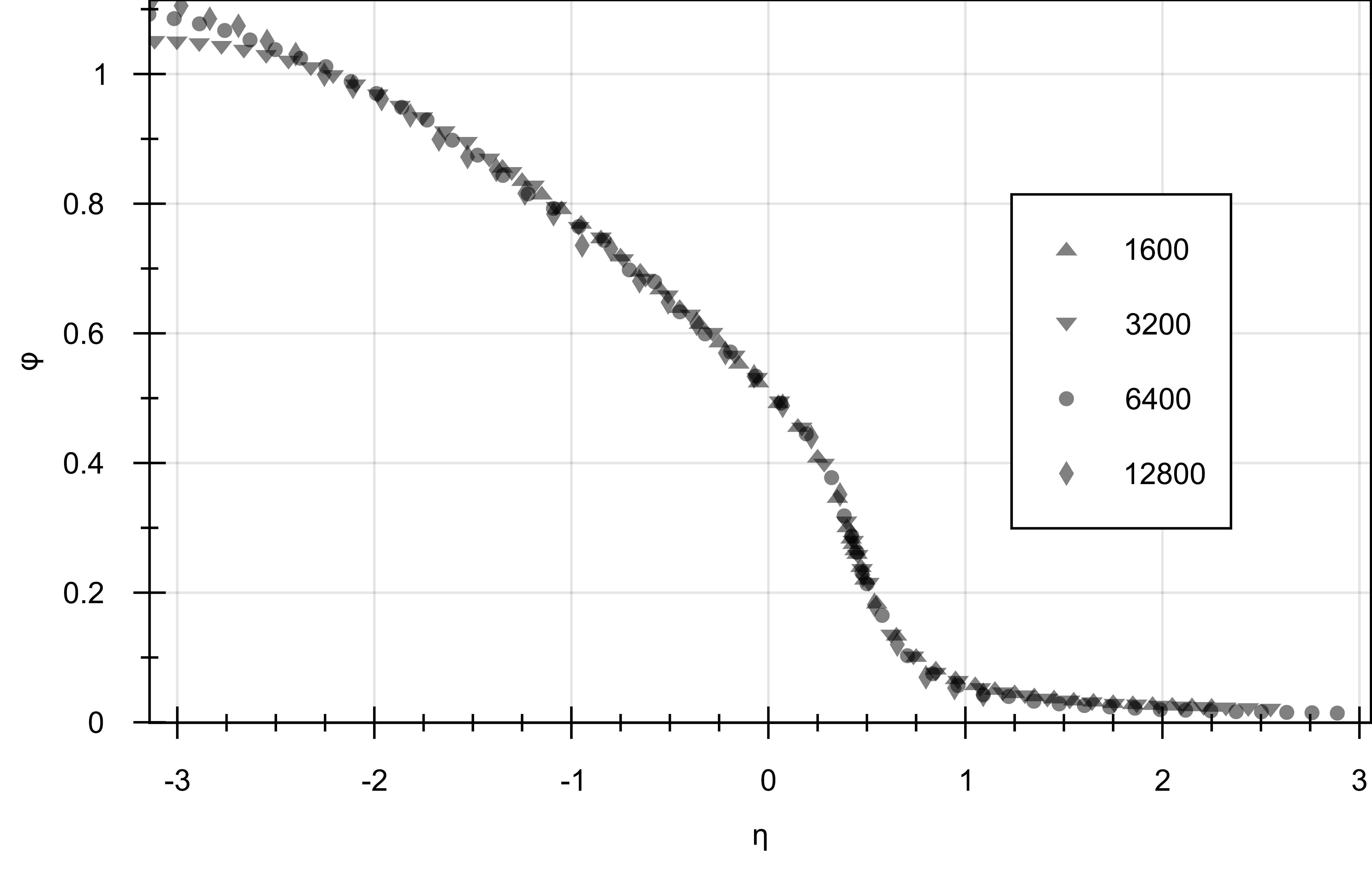}
\includegraphics[width=.45\textwidth]{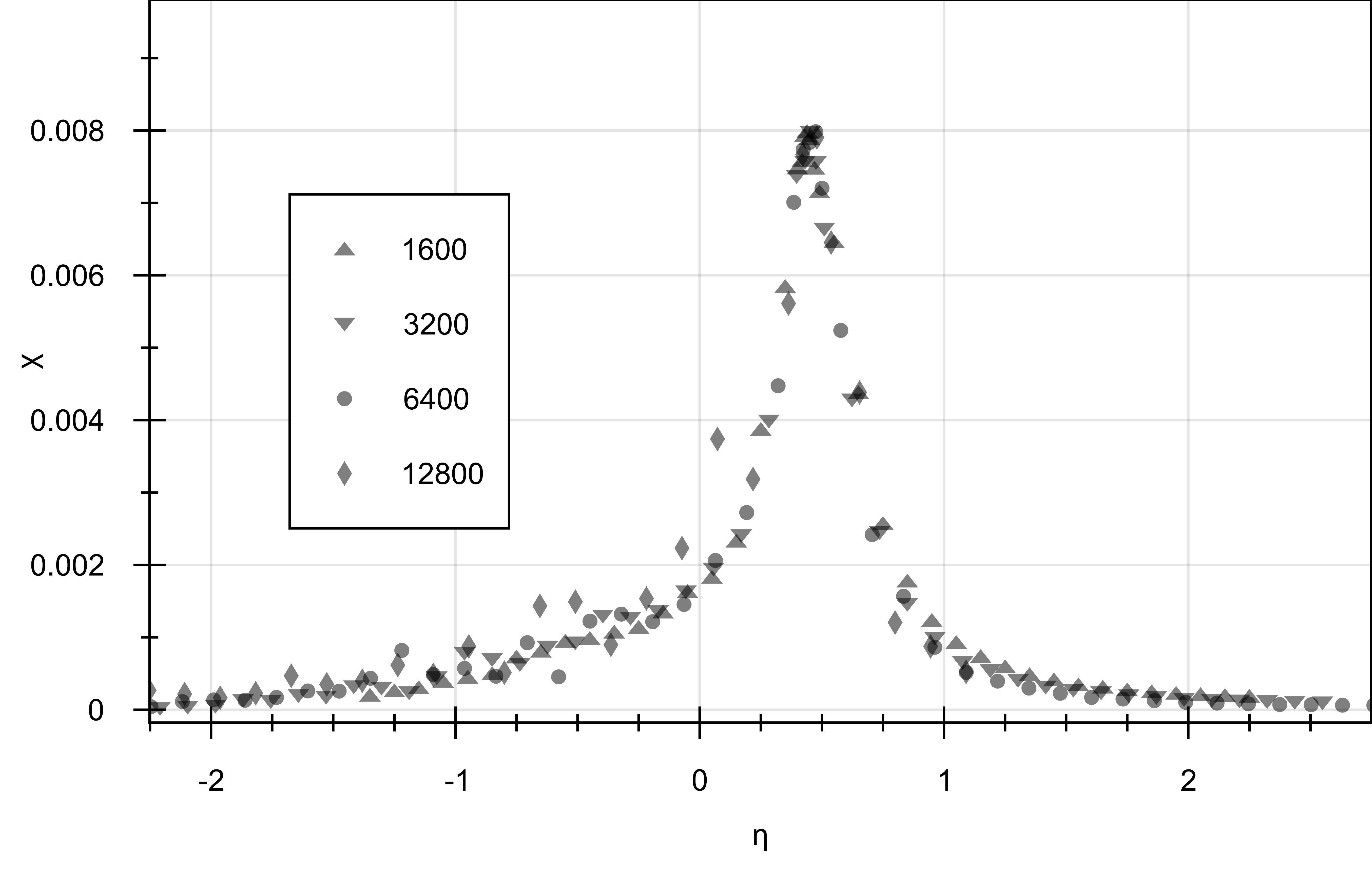}
\caption{The measured values of $ \varphi(\eta) $ and $ \chi(\eta) $ for a variety of $ N $ are rescaled to fit the $ N $-independent scaling functions.
\label{figscaling}}
\end{figure*}
The flipping algorithm also provides a partial solution to the task of updating the neighborhoods over time. After the individuals update their positions, as long as $ v $ is small with respect to the density, the edges from the previous time step may still form a graph which is in some sense close to being Delaunay. And there has been evidence that applying the flipping algorithm to a triangulation that is nearly Delaunay may be many times faster than building the Delaunay triangulation from scratch \cite{guibas}. However the flipping algorithm fails when the individuals move enough that the edges of the graph overlap (i.e. it is no longer a valid triangulation). We might simply apply the flipping algorithm assuming that the triangulation is valid, and to rebuild from scratch whenever an exception is encountered. For small $ N $ and $ v $ this may indeed perform better than simply rebuilding at each step. But the system sizes considered here are large enough that there is almost always some location where the edges of the graph overlap.

To deal with these overlapping edges, we first apply the ordinary flipping algorithm to the rest of the diagram--- only repairing the defective locations afterwards. The algorithm for the repair is inspired by the Star Splaying algorithm \cite{starsplay} with modifications to integrate it with the flipping algorithm and the particular data structure used (See Appendix \ref{repair}).

\section{Finite-Size Scaling}
\label{fss}

As was originally reported for the metric Vicsek model \cite{vicsek}, the density-independent model exhibits a continuous phase transition as described by Eq.\eqref{beta}. To better compare the new model with the Vicsek model, extensive simulations were carried out and a set of critical exponents was calculated. The method to determine the exponents closely follows the finite-size scaling analysis of the Vicsek model by Baglietto and Albano \cite{albano08}. Using the length scale determined by setting $ \rho = 1 $, a similar order of magnitude of speed $ v = .01 $ was chosen.

The simple power law behavior through which we defined the critical exponents in the thermodynamic limit is obscured by finite-size effects. We instead treat the system size $ N $ as an additional parameter of the system, and investigate the scaling behavior of the order parameter $ \varphi(\eta,N) $ and the susceptibility $ \chi(\eta,N) $.
\begin{equation}
\varphi(\eta,N) = N^{-\beta/2\nu}\tilde{\varphi}((\eta-\eta_C) N^{1/2\nu})
\label{fssphi}
\end{equation}
\begin{equation}
\chi(\eta,N) = N^{\gamma/2\nu}\tilde{\chi}((\eta-\eta_C) N^{1/2\nu})
\label{fsschi}
\end{equation}

Here $ \tilde{\varphi} $ and $ \tilde{\chi} $ are scaling functions which are independent of $ N $.

To help determine $ \eta_C $ we will make use of the Binder cumulant $ U $, an observable which tends to sharply distinguish the ordered and disordered phases. 
\begin{equation}
U=1-\frac{\langle \phi^4 \rangle}{3 \langle \phi^2\rangle^2}
\label{binder}
\end{equation}
The Binder cumulant has the useful property that the value of $ U(\eta_C) $ depends only weakly on $ N $. So the location of $ \eta_C $ can be determined from the point at which the curves $ U(N,\eta) $ intersect for various values of $ N $. By collecting data for three sizes in Fig.\ref{figbinder}, we took the intersection point to be $ \eta_C = 2.75 $.

Using Eq.\eqref{fssphi} and our value of $ \eta_C $, we can determine $ \beta / 2\nu $ from the slope of the line of best fit to the measured values of $ \varphi(\eta_C,N) $. A value of $ \beta / 2\nu = .069 \pm .002 $ was calculated, where the error was rescaled after considering the goodness of fit to the regression.

Similarly, we may determine $ \gamma/2\nu $ from Eq.\eqref{fsschi} and the values of the local maximum of the susceptibility $ \chi(\eta_{\text{Max}},N) $. In this manner we calculate a value of $ \gamma/2\nu = 0.866 \pm .001 $. These combinations of critical exponents are inconsistent with those previously calculated for the Vicsek model \cite{albano08}, indicating that this density-independent model is in a new universality class. However, these exponents also satisfy the hyperscaling relationship in Eq.\eqref{hyperscaling} within the bounds of the statistical error:
\begin{equation*}
1-2\frac{\beta}{2\nu} -\frac{\gamma}{2\nu}=-.0075\pm.0083\approx 0
\end{equation*}

The determination of $ 1/2\nu $, and thus $ \beta$, $\gamma,$ and $\nu $ alone, is more uncertain. Continuing to follow the approach in \cite{albano08}, we can determine $ 1/2\nu $ from the positions $ \eta_{\text{Max}} $ of the local maximum of $ \chi $. From Eq.\eqref{fsschi} we find that $ \eta_{\text{Max}}(N) $ should scale as $ \eta_{\text{Max}}-\eta_C \sim N^{-1/2\nu} $. With this method we calculate $ 1/2\nu = .13 \pm .02 $. However, this value leads to a poor overall fit to Eq.\eqref{beta}, \eqref{fssphi}, and \eqref{fsschi}.

Instead we consider the position of the inflection point of $ \varphi $, which should follow the same scaling law as $ \eta_{\text{Max}} $--- indeed, the two positions happen to be very close. With this method we instead calculate $ 1/2\nu = .20 \pm .04 $. Using this value and the previous two combinations of critical exponents, we can collapse the measured data onto the scaling functions $ \tilde{\varphi} $ and $ \tilde{\chi} $, as shown in Fig.\ref{figscaling}.

Calculating the order parameter critical exponent from these values, we find $ \beta = .34 \pm .08 $. We may instead calculate $ \beta $ directly from the function $ \varphi $ as in \cite{vicsek}. We find for $ N = 12800 $, $ \varphi $ fits the ansatz of Eq.\eqref{beta} well in the range $ .9 \leq \eta \leq 2.1 $, giving $\beta = .35 \pm .01 $. If we accept the validity of this determination of $\beta$, we can further calculate $ \gamma = 4.4 \pm .2 $ and $ \nu = 2.5 \pm .1 $. If we instead use the less precise measurement of $1/2\nu $ to calculate the critical exponents, the error increases by a factor of 5.

\begin{figure}[!b]
\includegraphics[width=.4\textwidth]{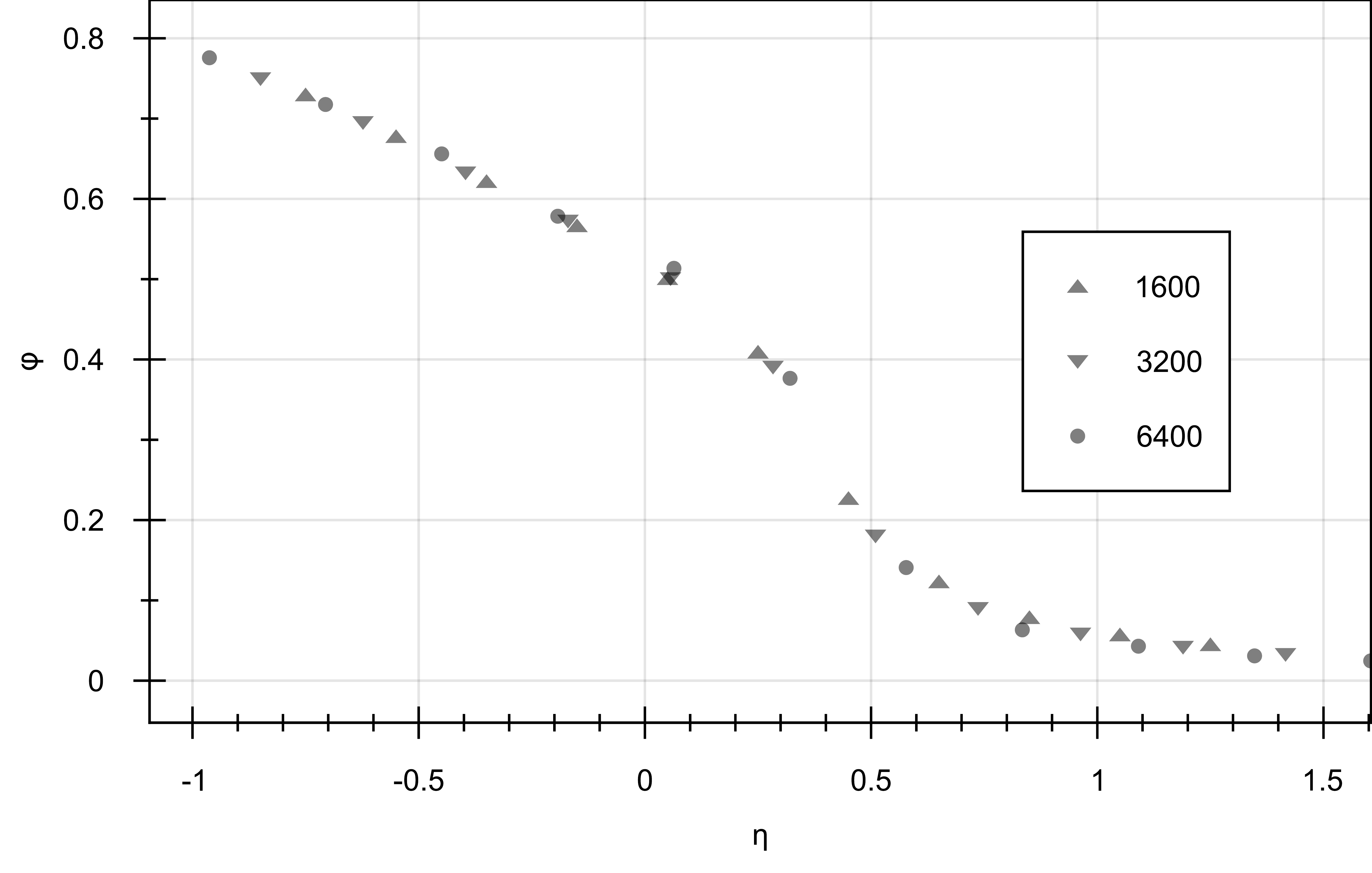}
\caption{The values of $ \varphi $ for different system sizes in the repulsive model ($ r_c= 0.1 $, $ \beta= 2.5 $) are fit to a scaling function using the same critical exponents as the model without repulsion.\label{figrepulse}}
\end{figure}
\section{ Repulsion}
\label{repulsion}

Having calculated a new set of critical exponents, we might wish to examine the scope of this universality class. One of the most common features of collective motion models is the existence of a repulsion term to maintain a safe distance around each individual \cite{review}. Also, in this density-independent model, localized `flocks' do not form in the absence of an explicit attractive force of some kind. So the adoption of some counterbalancing repulsion term is a prerequisite for a full flocking model based on Voronoi neighborhoods.

To incorporate repulsion, we add an additional term in the calculation of the average direction $ \langle\theta(t)\rangle_{N_i} $ in Eq.\eqref{velupdate} 
\begin{equation}
\langle\theta(t)\rangle_{N_i} = \arg\Big [ \sum_{N_i}\mathbf{v} + f(\|\mathbf{e}\|) \hat{\mathbf{e}}\Big ].
\end{equation}
In this expression $ \mathbf{e} $ is the edge pointing from the individual $ i $ to one of its neighbors, and the repulsion force magnitude $ f $ is given as
\begin{equation}
f(d)=\frac{\beta}{1-e^\frac{d}{r_c}}
\end{equation}
Following the naming conventions of an earlier cohesive model \cite{gregoire03}, $ r_c $ is a parameter controlling the range of the force, and $ \beta $ (not to be confused with the critical exponent) controls the strength of the force with respect to the velocity matching behavior.

As expected, this repulsion term limits the local density fluctuations in the model, and the dynamics may at first appear qualitatively different. Even so, the model still appears to obey a phase transition with the same critical exponents.

Rather than recalculating the critical exponents as in Section \ref{fss}, we use the existing critical exponents to fit the measured values of the repulsive model to $ \tilde{\varphi} $ using  Eq.\eqref{fssphi}. In Fig.\ref{figrepulse} we plot the data for $ r_c= 0.1 $ and $ \beta= 2.5 $, also taking the critical point location to be the same as that of original model. The shape of the scaling function $ \tilde{\varphi} $ is slightly different from the case with no repulsion (and the case with $ \beta = 1.0 $, also considered), but the presence of the repulsion term does not appear to lead to a different universality class.

\begin{figure}[!h]
\includegraphics[width=.4\textwidth]{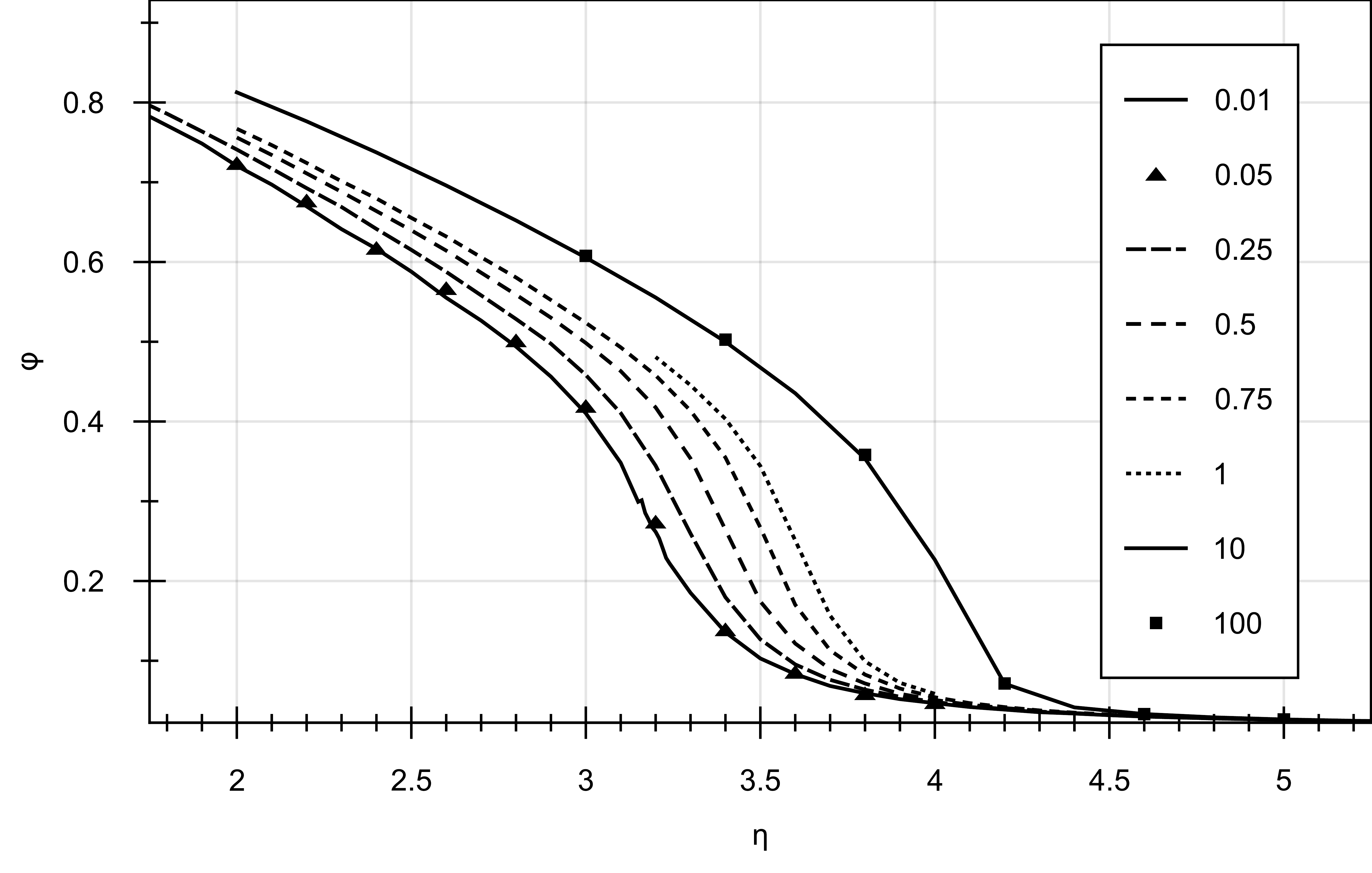}
\caption{A plot of $ \varphi $ at $ N = 1600 $ for various $ v $.\label{figvel}}
\end{figure}

\section{High-Velocity Regime}

The critical exponents of Section \ref{fss} and \ref{repulsion} were calculated with the speed parameter fixed at $ v = .01 $. Following many authors on the Vicsek model, we have so far limited our discussion to what has been called the low-velocity regime \cite{aldana}. We have already mentioned in Section \ref{delaunay} that a small value of $ v $ allows us to use the algorithm given in Appendix \ref{repair} to significantly reduce the processing time per step. But we have not yet considered its role in the physics of the model.

\begin{figure*}[!t]
\includegraphics[width=.4\textwidth]{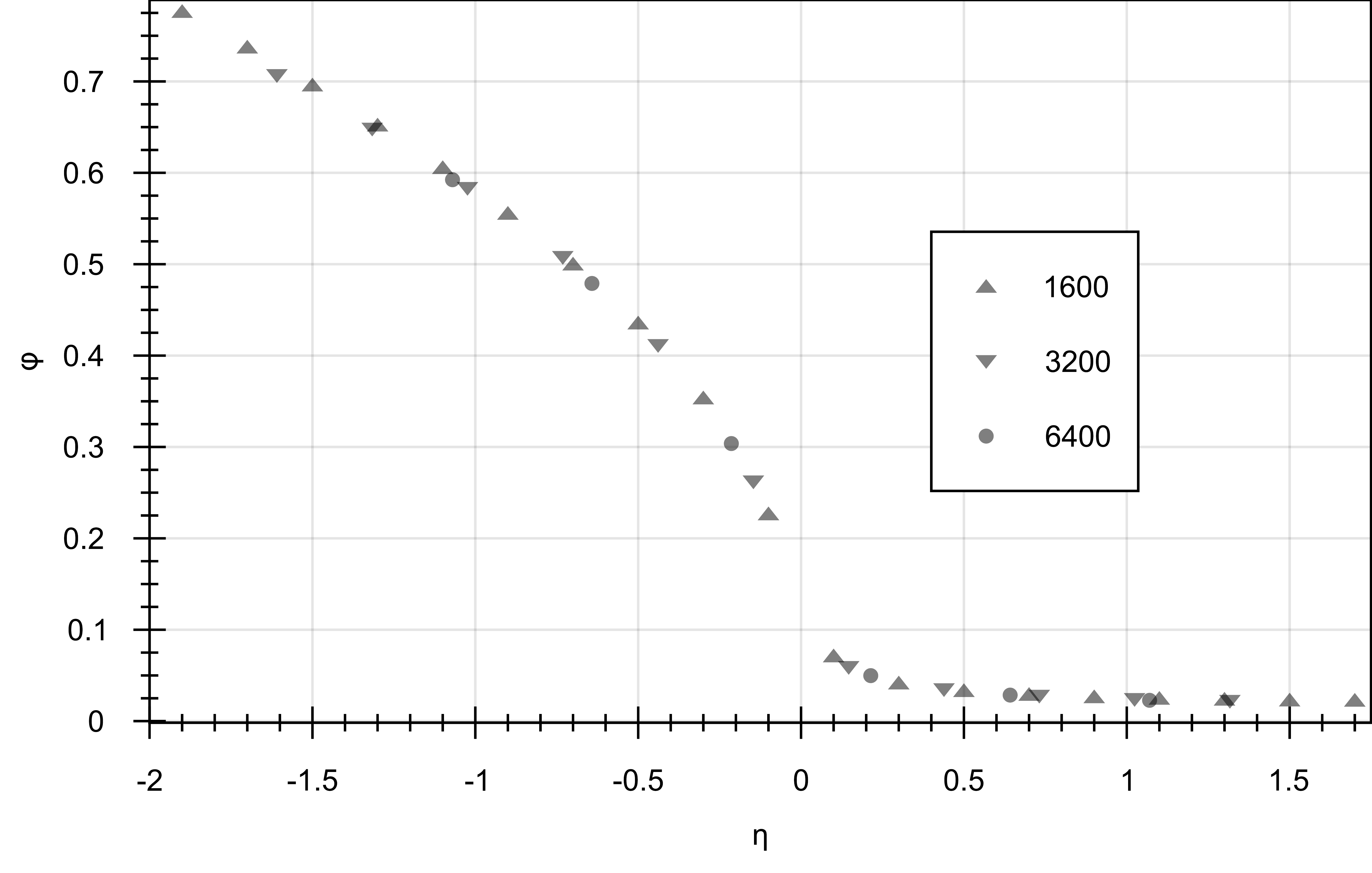}
\includegraphics[width=.4\textwidth]{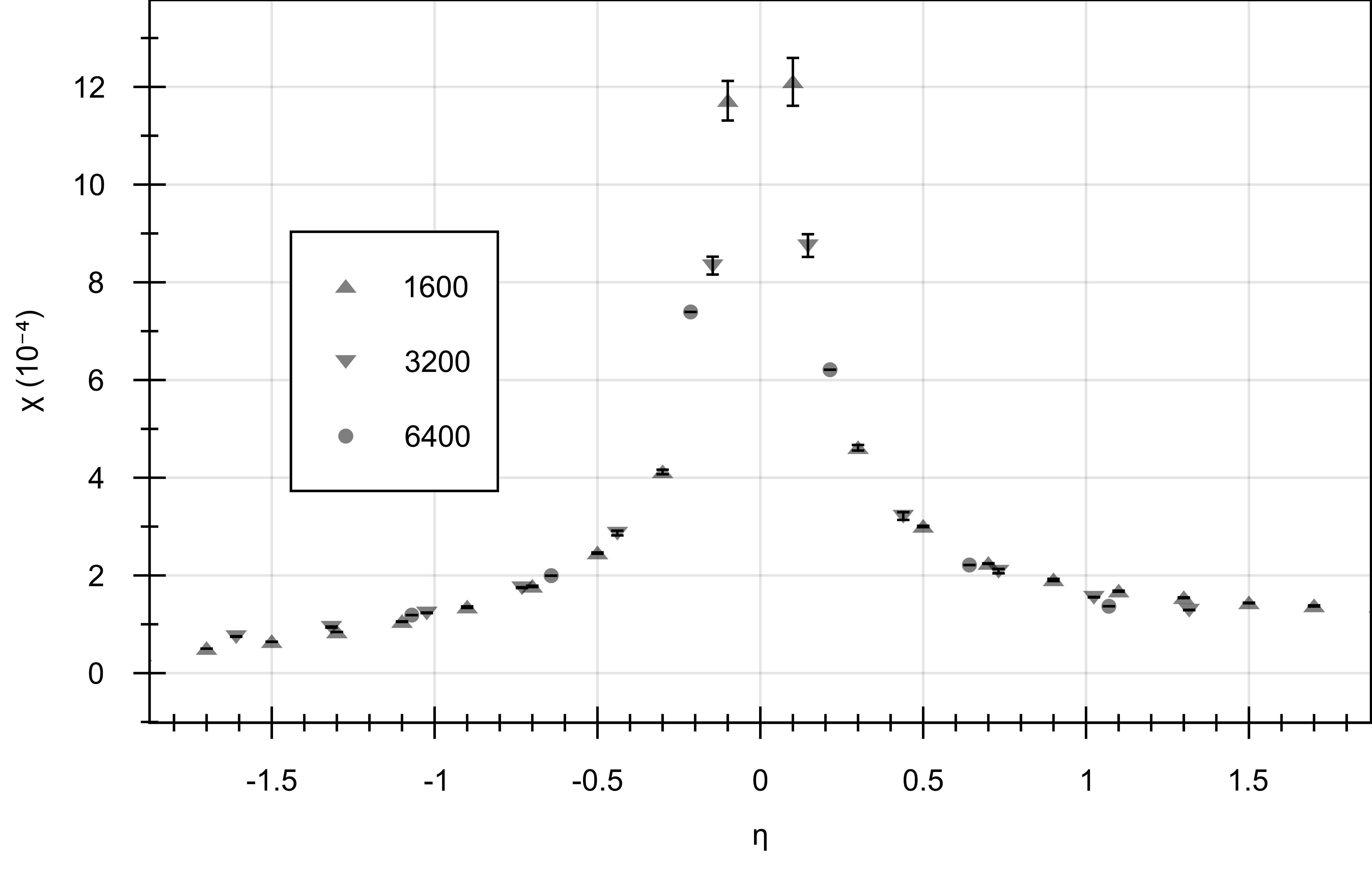}
\caption{The values of $ \varphi(\eta) $ and $ \chi(\eta) $ for $ v = 10 $ fit to the scaling functions using the high-velocity regime critical exponents.
\label{fighv}}
\end{figure*}

Already in the early papers on the Vicsek model there are claims that the choice of the parameter $ v $ is not important, at least within some low-velocity regime \cite{vicsek,vicsek2}. In the density-independent model we too find that for very small $ v $ the physics seems robust (see Fig.\ref{figvel} for $ v = .01 $ and .05). As $ v $ increases, the critical point begins to shift until we reach a high-velocity regime which also seems fairly insensitive to changes in $ v $ (here consider $ v = 10 $ and $ v = 100 $).  Preliminary investigations suggest that this high-velocity regime is distinct from a model with random position reassignment at each time step.

It may not be surprising that the shape of the scaling functions and the position of the critical point depend on $ v $, but we also find that the critical exponents themselves change with $ v $. To demonstrate this we will calculate a new set of critical exponents in the high-velocity regime.

As can be observed from Fig.\ref{figvel}, the high-velocity regime seems to fit the scaling ansatz given by Eq.\eqref{beta} with weaker finite-size effects than the low-velocity regime. As such, we can follow the original papers on the Vicsek model, and use this scaling law directly to find $ \beta $, with $ \eta_C $ determined so that the data best fits the power law \cite{vicsek,vicsek2}. We then assume that the hyperscaling relationship in Eq.\eqref{hyperscaling} holds, which leaves only one remaining independent critical exponent, which we adjust to give the best fit to the scaling functions in Eq.\eqref{fssphi} and \eqref{fsschi}.

In Fig.\ref{fighv} we see that the scaling laws appear to hold when we choose  $ \eta_C = 4.1,\,\beta=.41,\,\nu=.91, $ and $ \gamma=1.0 $. Tests at the intermediate value of $ v = 0.5 $ suggest that the critical exponents vary continuously between the low and high velocity regimes, although the values determined by fitting to Eq.\eqref{beta} were less clear in this case. Interestingly, the model of Ginelli and Chat\'{e} was conducted at $ v = 0.5 $, and their critical exponents $ \beta/\nu = .23(3) $ and $ \gamma/\nu = 1.49(5) $ are intermediate between the critical exponents for the low and high velocity regimes calculated here \cite{ginelli}. However, their value of $ \nu $ was found to be less than that caluculated in either limit. Certainly the velocity dependence of the critical exponents merits further study.

\section{Correlation Lengths}

One of the claims made by the team investigating starling flocks was that the velocity correlation length is always proportional to the linear size of the flock \cite{correlations}. One of the motivations behind developing the density-independent Vicsek model was to study whether the observed scale-free correlations were in some way related to an interaction based on topological distance. It is important to note that for flocks with the same number of individuals but different densities, the scaling behavior follows trivially in the density-independent model. A previous model based on a different implementation of topological distance has also exhibited this correlation length scaling for flocks of the same numerical size \cite{starmodel}.

\begin{figure}[!b]
\includegraphics[width=.4\textwidth]{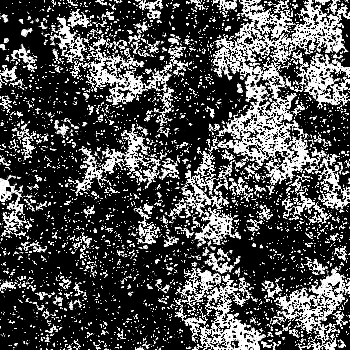}
\caption{Sample view of the density-independent model at the critical point ($ v = .01 $, $N = 4.096\times 10^5$, $\eta = 2.75$ ).  The Voronoi tesselation is shown, and a cell is shaded black or white depending on the magnitude of the velocity deviation.
\label{figcrit}}
\end{figure}
\begin{figure*}[!t]
\includegraphics[width=.4\textwidth]{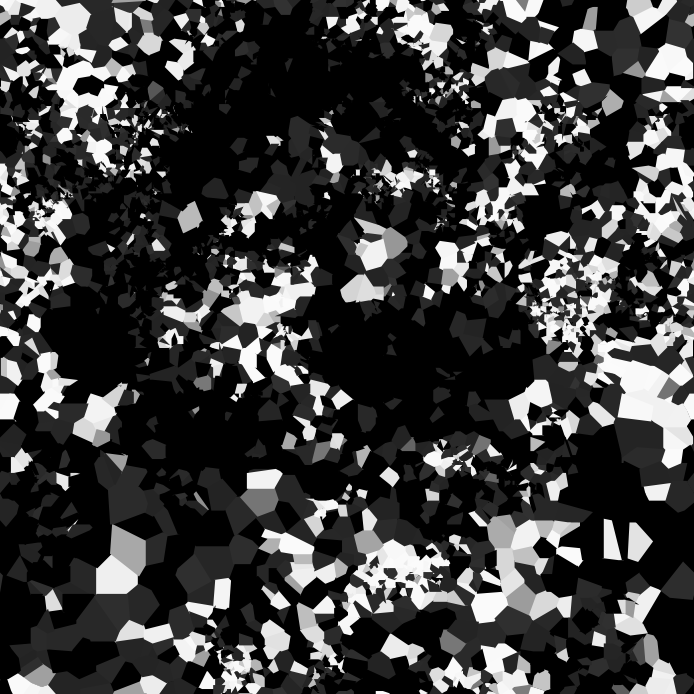}
\includegraphics[width=.4\textwidth]{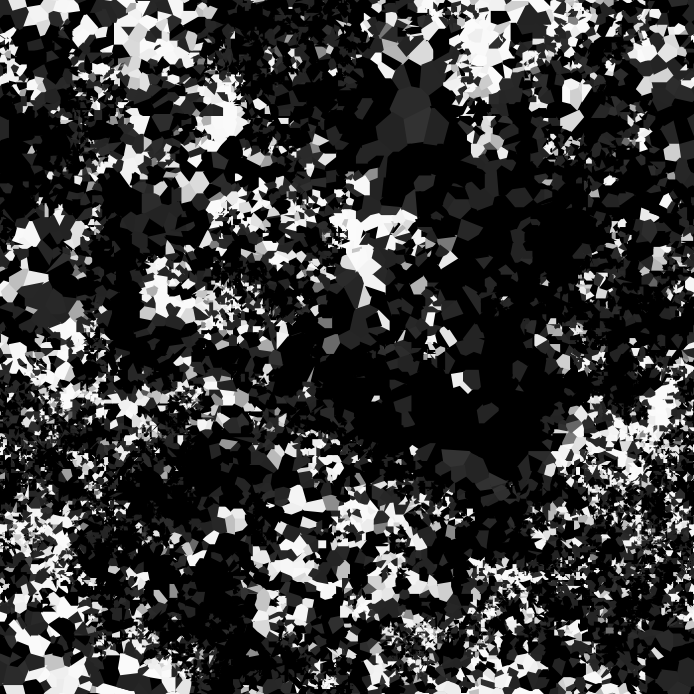}
\caption{Sample Voronoi tesselations for $ N = 6,400 $ and $ N = 25,600$ far from the critical point ($ v = .1 $, $ \eta = 1.1$). The order parameter $ \phi > .9 $ in both cases. Cells with velocity deviations to one side of the average are shaded black, while those on the other side are shaded with a brightness depending on the magnitude of deviation.
\label{figcritln}}
\end{figure*}

\begin{figure}[!b]
\includegraphics[width=.4\textwidth]{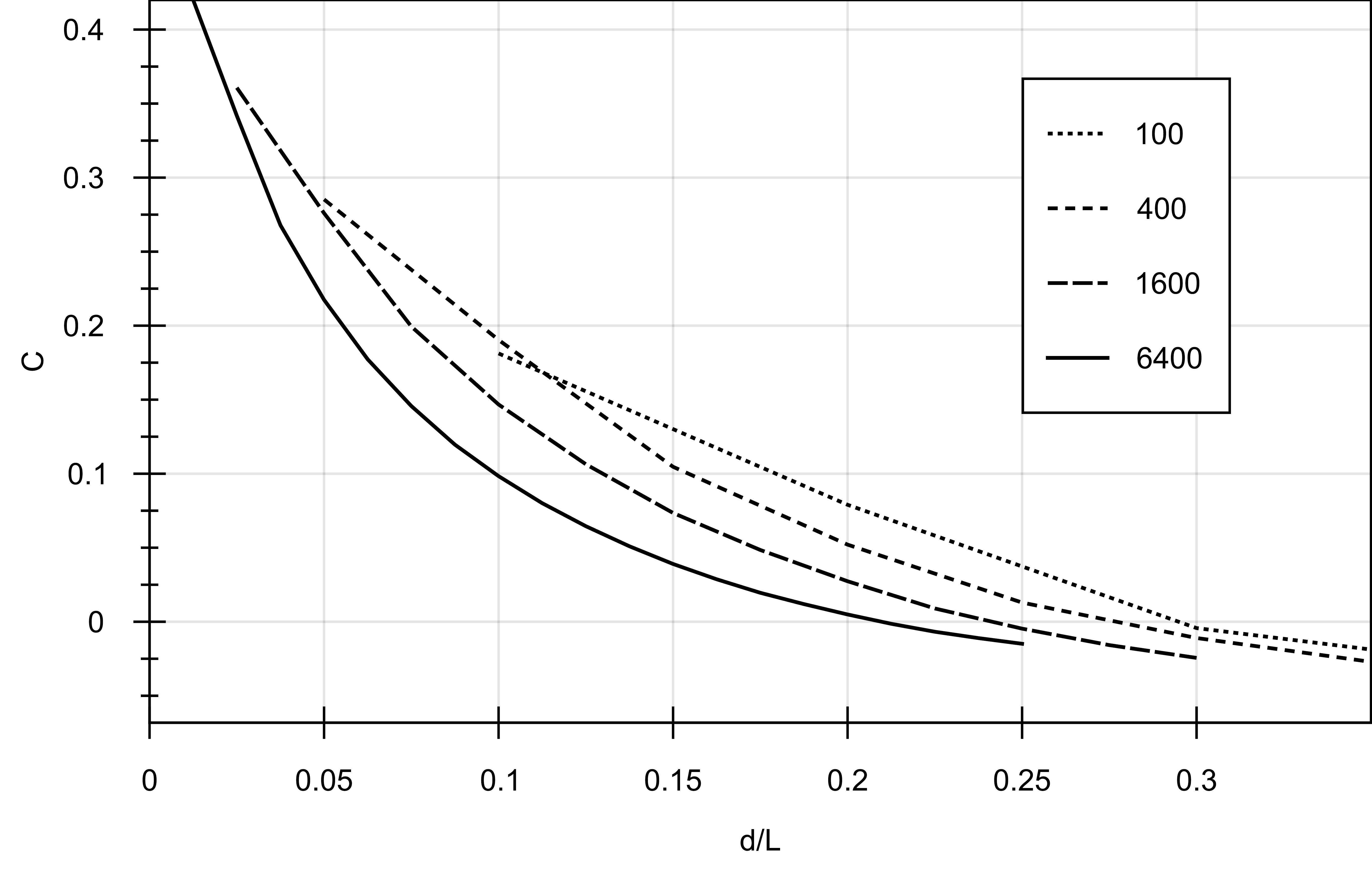}
\caption{The normalized graph distance correlation functions $ C $ for systems of varying size far from the critical point ($ v= .1 $, $ \eta = 1.1 $). The horizontal axis represents the ratio of the graph distance $ d $ to the length scale $ L $ \label{figcorr}}
\end{figure}

To test whether a model of collective motion fully exhibits scale-free correlations we need to test whether the correlation length scales appropriately for flocks of different numerical sizes. The observational group did indeed analyze flock sizes ranging from 122 to 4,268 \cite{correlations}. Their hypothesis was that the correlation length would diverge in the thermodynamic limit - suggesting the system is at a self-organized critical point. But it is important to note that a system with a large but finite correlation length may also exhibit scaling of correlation length over small system sizes due to finite-size effects.

With systems near the critical point, we would of course expect the scale-free correlations to be observed. A Voronoi tesselation with cells shaded depending on  indivual velocity deviations can reveal fluctuations on the order of the system size (Fig.\ref{figcrit}). The crucial difference in the starling research is that the dynamics do not appear to be tuned to a special `noise' critical point. In particular, the average order parameter of the flocks was as high as $ 0.96 \pm 0.03 $ \cite{correlations}.

To test whether the scaling behavior holds in the ordered phase, we compared a form of correlation length for a variety of system sizes. Rather than considering the correlations in terms of metric distance, we use the preexisting Delaunay triangulation to find the correlation over graph distance. While there will be discrepancies for small graph distances (consider the regular hexagonal Delaunay triangulation, for example), we assume that the graph correlation length is assymptotically proportional to the metric correlation length.

We calculated the correlation functions (normalized to one at zero distance) for systems at $ v = .1, \eta = 1.1 $. The order parameter was above .9 for all system sizes. In Fig.\ref{figcorr} we plot these functions after recaling the graph distance by the system length $ L = \sqrt{N} $.  If we define the correlation length as the point where the correlation function first equals zero as in \cite{correlations}, we observe that the rescaled correlation length decreases only slowly with $ N $. For comparison, a similar small $ N $ dependence was observed for systems at the critical point itself. This is perhaps due to the fact that for larger system sizes there are proportionately smaller length scales which are suppressed by the lattice spacing in smaller systems. If this is the case, the scaling should become more exact for even larger sizes.

Considering that the starling research analyzed only a small number of flocks ---some of which may vary mostly in density rather than number--- this weak decrease in correlation length for small $ N $ may explain the observation of apparent scale-free correlations among starling flocks. The apparent scale-free fluctuations in the density-independent model are visualized using the Voronoi tesselation in Fig.\ref{figcritln}. Preliminary investigations suggest that these results also hold for even smaller noises, but that the correlation length is indeed finite and decreasing above the critical point.

These results do not necessarily suggest self-organized criticality in the density-independent model. It is possible instead that the correlation length follows the expected power law behavior described by the critical exponent $ \nu $ near and above $ \eta_C $, but also diverges as $ \eta \rightarrow 0 $, maintaining a large value throughout the ordered phase. Further research is needed to test these possibilities.

\appendix

\section{Triangulation over periodic boundary conditions}
\label{boundary}

We first construct any triangulation bounded by the convex hull. We keep a list of the points on the convex hull, and orient them counterclockwise.

We consider the space as a periodic Euclidean plane, and will connect the tiled copies of the convex hull together. In particular we will consider two copies, one on the left and the right, and we will first connect these copies to those above and below.

\begin{figure}[!hb]
\includegraphics[width=.45\textwidth]{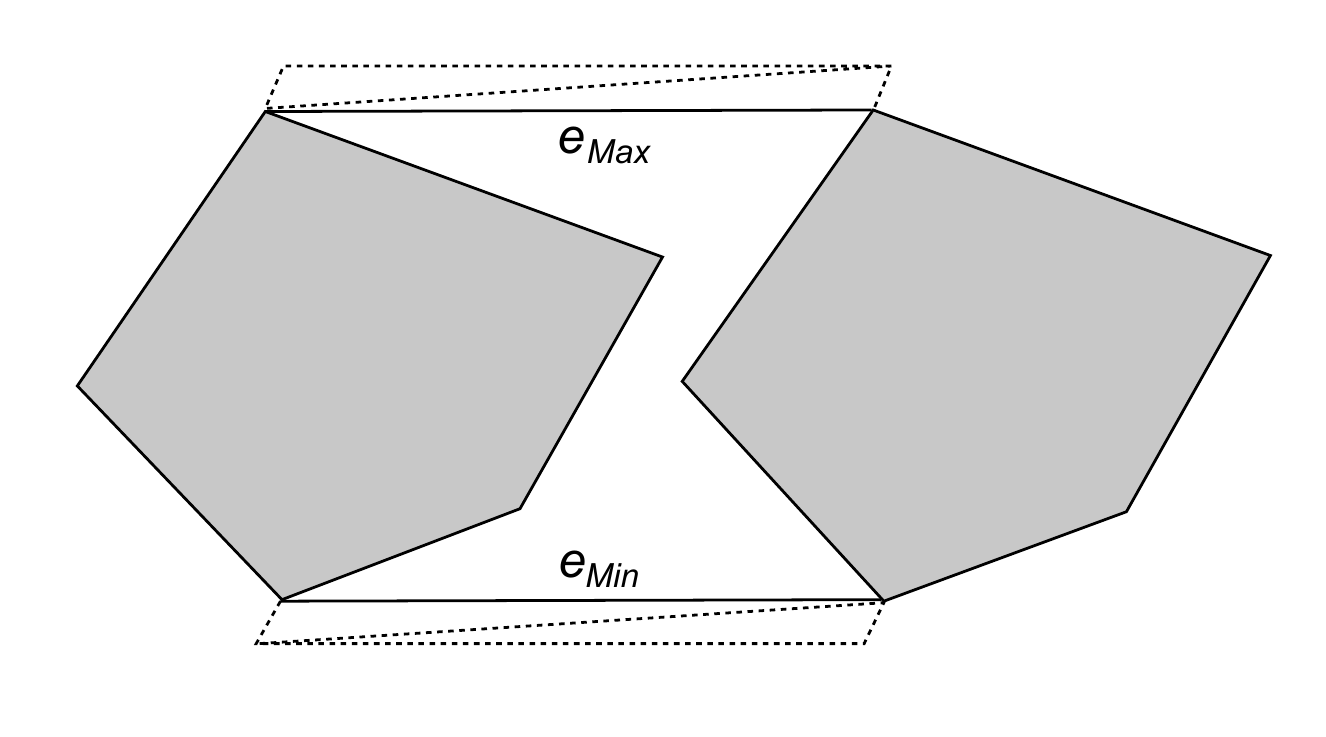}
\caption{Connecting two periodic copies of the convex hull at the points $ p_{\text{Max}} $  and  $ p_{\text{Min}} $\label{figa11}}
\end{figure}

We find  $ p_{\text{Max}} $  and  $ p_{\text{Min}} $ , the points with greatest and lowest $ y $ value on hull. We form the edges $ e_{\text{Max}} $  and  $ e_{\text{Min}} $ connecting $ p_{\text{Max}} $  and  $ p_{\text{Min}} $ to themselves, and also two diagonal edges connecting $ e_{\text{Max}} $  and  $ e_{\text{Min}} $, which wrap around the space in different ways (See Fig.\ref{figa11}). These edges are degenerate in the sense discussed above, and will require special treatment in the flipping algorithm.

\begin{figure}[!ht]
\includegraphics[width=.45\textwidth]{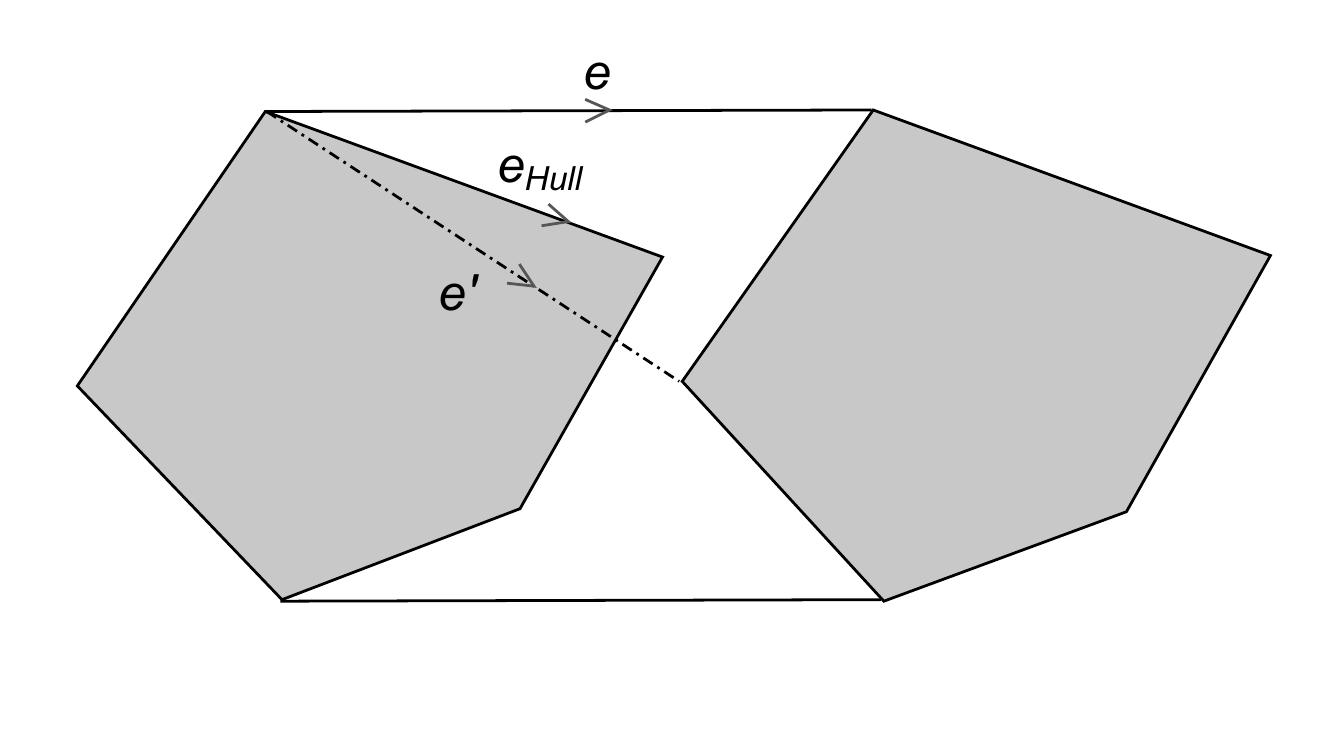}
\caption{If $ e_{\text{Hull}} \times e'$ points into the page, the test edge $ e' $ is invalid\label{figa12}}
\end{figure}

Next we will stitch together the hull between $ e_{\text{Max}} $  and  $ e_{\text{Min}} $. We start with $ e = e_{\text{Max}} $ and construct a new edge each step, moving an endpoint on either the left or right copy down along the hull until we reach $ e_{\text{Min}} $.
First we try to move the right side endpoint down, forming the test edge $ e' $ (Fig.\ref{figa12}). We will also consider the edge along the left side hull $ e_{\text{Hull}} $. The edges $ e $ and $ e’ $ are always oriented pointing from the left copy to the right. The edge $ e_{\text{Hull}} $ is oriented pointing clockwise around the hull. If $ e_{\text{Hull}} \times e' < 0$, the test edge will pass through the left side hull and can't be valid (invalid edges are here plotted with a dash-dot pattern). We construct the edge by moving the left side instead.

\begin{figure}[!ht]
\includegraphics[width=.45\textwidth]{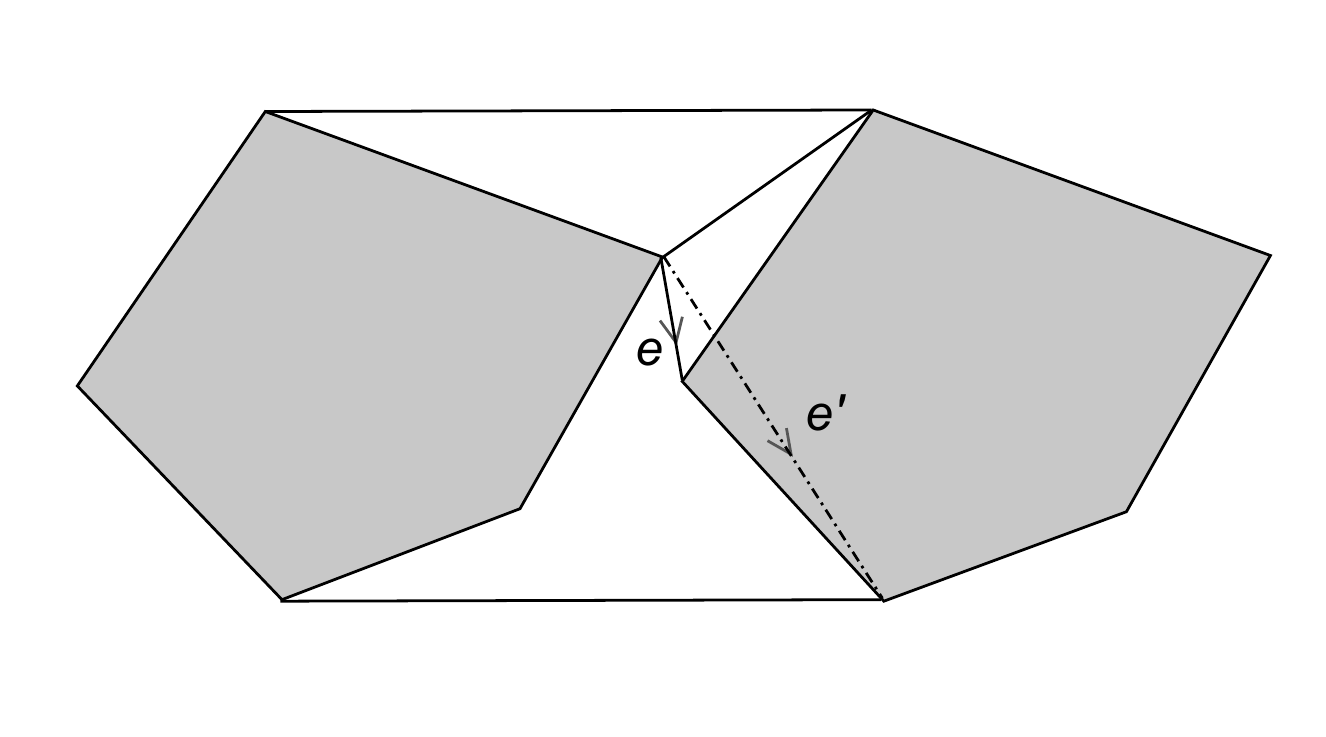}
\caption{Similarly if $ e' \times e$ points into the page, the test edge will pass through the right side hull\label{figa13}}
\end{figure}

\begin{figure}[!ht]
\includegraphics[width=.45\textwidth]{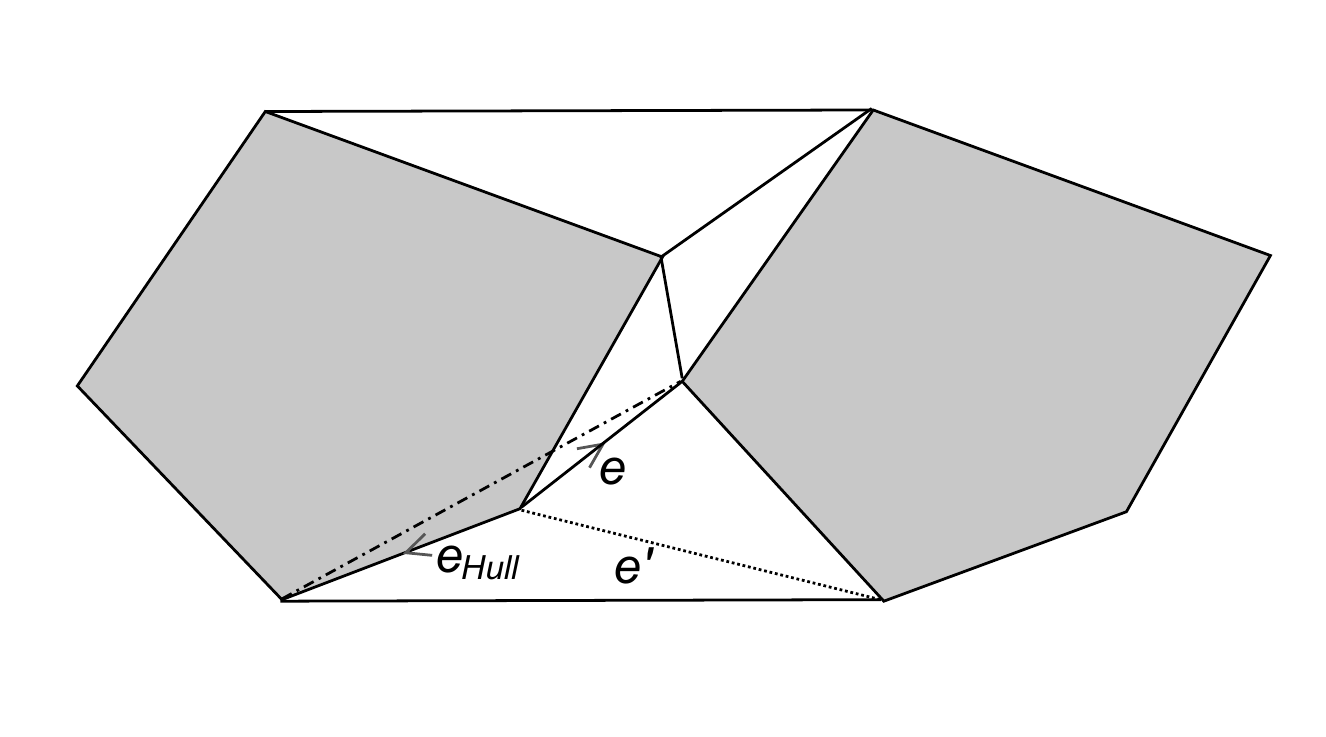}
\caption{Finally, if $ e_{\text{Hull}} \times e$ points into the page, we are guaranteed to have $ e' $ be valid. This is because the edge we would construct on the left side would pass through the hull.\label{figa14}}
\end{figure}

In the actual algorithm, we first test the condition represented by Fig.\ref{figa13}, then if that passes we test the condition represented by Fig.\ref{figa14}. If it passes that test we finally check the condition of Fig.\ref{figa12}. Here the algorithm was explained out of order so that the conditions are encountered as the diagram for the example shape is constructed.

\section{Repairing the Delaunay triangulation}
\label{repair}
\begin{figure}[!ht]
\includegraphics[width=.2\textwidth]{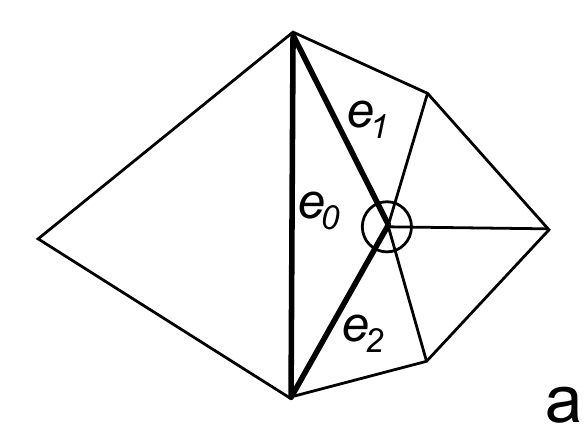}
\includegraphics[width=.2\textwidth]{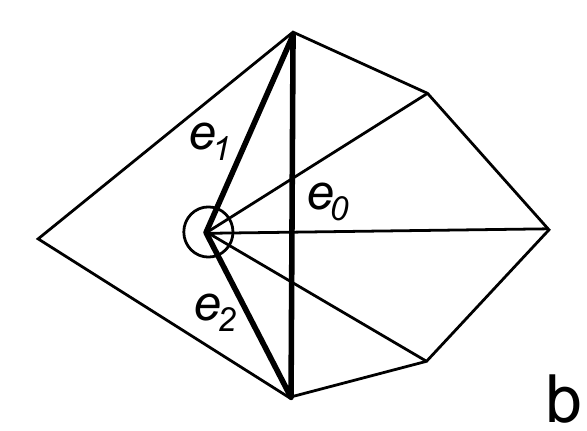}
\caption{The circled point $ P $ and its neighboring edges move from the initial position (a) to the position (b) with invalid overlapping edges. \label{figa21}}
\end{figure}

As the individuals move, the graph defining the neighbors may not only fail to have the Delaunay property--- it may also fail to be a valid triangulation. This occurs when an individual crosses one of the edges of the graph (consider the point $ P $ labeled by a circle in Fig.\ref{figa21}).

Each edge of $ P $ (and every other individual) contains an address to the nearest neighboring edge on both the left and right. In a valid triangulation, this nearest neighbor information allows us to traverse the edges surrounding $ P $ in a strictly counter-clockwise orientation. But when the triangulation is broken as in Fig.\ref{figa21}b, the nearest neighbor information from the previous time step leads to changes in orientation which we can detect by evaluating cross-products.

\begin{figure}[!ht]
\includegraphics[width=.2\textwidth]{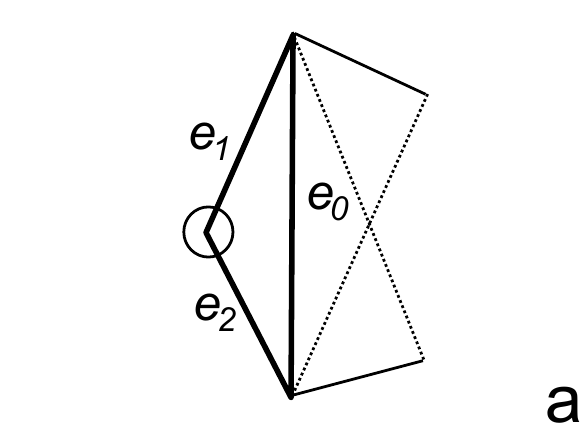}
\includegraphics[width=.2\textwidth]{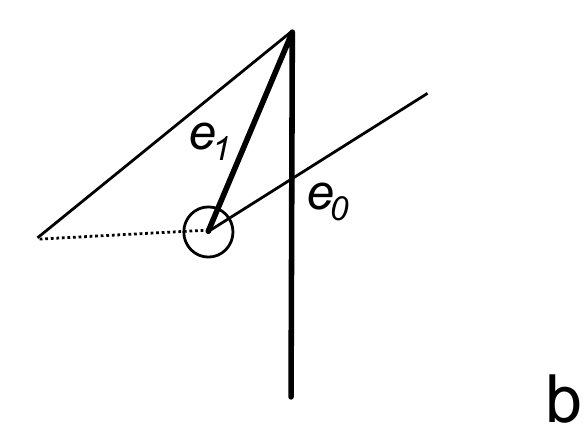}
\caption{After updating the nearest neighboring edges, an edge may no longer have two valid neighboring triangles. (a) The neighborhood of the edge $ e_0 $. The right side can be repaired by adding either of the two edges indicated by a dotted line. (b) The neighborhood of $ e_1 $. There is only one neighboring edge on the left side, which can be repaired by adding the unique dotted line. \label{figa22}}
\end{figure}

In particular, note that when two edges of a triangle have flipped orientation - for instance the triangle in bold formed by $ e_0 $, $ e_1 $, and $ e_2 $ - the third edge must also flip orientation with the other two edges. So not only $ P $, but also the other two vertices of the triangle have defective edge neighborhoods. As a first step to repairing the triangulation we detect all points with defective neighborhoods and update the nearest neighbor information. Note that in Fig.\ref{figa21}b, $ e_1 $ and $ e_2 $ will no longer be considered neighbors since there is a counter-clockwise angle of greater than $ 180^\circ $ between them. After this update, the four (or less) neighboring edges of some of the corrected edges may no longer form a valid quadrilateral (see Fig.\ref{figa22}). For those edges that do have a valid quadrilateral neighborhood, we apply the usual Delaunay flipping algorithm. (Note that there is the possibility that the neighboring edges may not themselves be neighbors at their common point. This is not considered a valid quadrilateral for the purposes of this algorithm.)

In Fig.\ref{figa22}a, only the four neighboring edges of the edge $ e_0 $ in Fig.\ref{figa21}b are shown. On the left side, $ e_1 $ and $ e_2 $ do indeed form a valid triangle. But on the right side, the two neighboring edges (plotted as solid lines) do not meet at a common point. We must add either of the two dotted lines to the diagram to correct this. We choose the dotted line which forms the triangle with the greater opposite angle. In this case, the choice does not matter, but this criterion allows us to choose the interior triangle in the case that one triangle is contained within the other.

In Fig.\ref{figa22}b, the neighboring edges of $ e_1 $ in Figure \ref{figa21}b are shown ($ e_2 $ has a similar neighborhood). In this case, since $ e_1 $ and $ e_2 $ are not considered neighbors, there is only one edge on the left side. We can immediately correct this defect by adding the dotted line to the diagram.

On the right side, not only do the neighboring edges not meet at a common point, they also cross. We can't correct this as with Fig.\ref{figa22}a since the possible dotted lines would still not be nearest neighbors of $ e_1 $.

\begin{figure}[!ht]
\includegraphics[width=.3\textwidth]{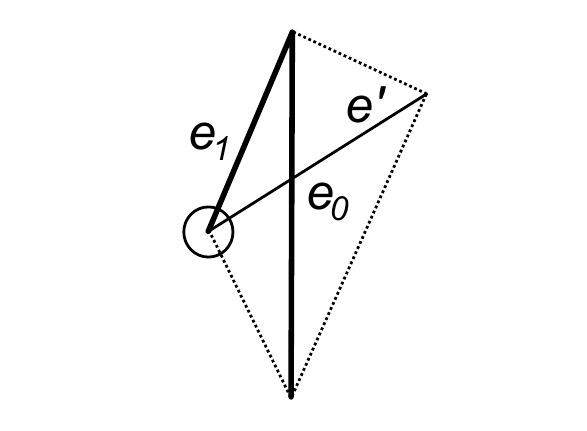}
\caption{The right side neighborhood of the edge $ e_1 $. Either $ e_0 $ or $ e' $ will fail the criterion to be a valid Delaunay edge and will be deleted. The dotted line edges are not added at this step since they may already exist in the diagram.\label{figa23}}
\end{figure}

Instead we consider the quadrilateral formed as in Fig.\ref{figa23}, with the thin solid edge now labeled $ e' $. We will delete either $ e_0 $ or $ e' $ depending on whether they satisfy the Delaunay criterion for this quadrilateral or not. We do not add any of the dotted lines to the diagram, since they may still not be nearest neighbors to $ e_1 $, or they may already exist in the diagram. Instead we simply reapply the algorithm to  $ e_1 $, and it will be corrected as one of the other defective cases.

Since in this case we are deleting an edge without adding a new one, in rare cases an edge will be left with no neighbors at all. In these cases an exception is thrown and the diagram is rebuilt from scratch. These exceptions only become a problem for velocities $ v $ high enough that it is more effective to rebuild every time step anyway. However it may be possible that a modification to the algorithm in the case of crossed edges will both fix this problem and lead to an increase in speed.

\section{Statistical analysis}
\label{stat}

Our analysis relies on calculating the ensemble average and higher order moments of the instantaneous order parameter $ \phi $. One possible approach to doing this might be to only record and consider $ \phi $ values separated by a large number of time steps $ T $ such that $ T $ is much greater than the correlation time $ \tau $. We instead make the assumption that $ \phi $ evolves ergodically--- So that by averaging the values of $ \phi $ and its moments at all time steps we may converge to the expected values faster. However, the calculation of errors in this approach is not as straightforward.

To analyze the data, we coarse-grain the time steps into cells of size $ T $. We call the coarse-grained variable $ X_1 $,
\begin{equation}
X_1(t) \equiv \langle\phi(t)\rangle_T \equiv \dfrac{1}{T}\sum^{t+T}_{k=t} \phi(k).
\label{coarse}
\end{equation}
\
Similarly, we define $ X_2 $ and $ X_4 $ by $ \langle\phi^2(t)\rangle_T $ and $ \langle\phi^4(t)\rangle_T $ respectively.

Note that the ensemble average $ \langle X_1\rangle = \langle \phi\rangle = \varphi $. Consider the variance of $ X_1 $ in terms of the correlation function $ R(i) = \dfrac{1}{\sigma^2} \langle(\phi(t)-\varphi)(\phi(t+i)-\varphi)\rangle $

\begin{eqnarray}
\text{Var}(X_1) & = & \langle(\dfrac{1}{T}\sum_i{\phi(i)-\varphi}) (\dfrac{1}{T}\sum_j{\phi(j)-\varphi})\rangle\nonumber\\
& = & T^{-2}(\sum_i\langle (\phi(i)-\varphi)^2 \rangle\nonumber\\
&&	+\:  \sum_{i\neq j}\langle(\phi(i)-\varphi)(\phi(j)-\varphi)\rangle)\nonumber\\
& = & T^{-2}(T \sigma^2 + 2 (T-1) \sigma^2 R(1) \nonumber\\
&& +\: 2 (T-2) \sigma^2 R(2) + \ldots)\nonumber\\
& = & \dfrac{\sigma^2}{T}(1+2\sum_i R(i))- \dfrac{\sigma^2}{T^2}\sum_i i R(i)\nonumber
\end{eqnarray}

For large $ T $, we may ignore the second term on the right-hand side and so the variance of the $ X_1 $

\begin{equation}
\text{Var}(X_1) \approx \dfrac{\sigma^2}{T}(1+2\sum_i R(i)).
\label{approxvar}
\end{equation}

Note that if $ T $ is large enough that \eqref{approxvar} holds, then the variance of an integer multiple coarse-grained steps $ k T $ is simply $ 1 / k \text{Var}(X_1) $. So we may treat the variables $ X_1 $ as if they were statistically independent. We have assumed further that the higher order momentsmay also be treated as independent for the same size $ T $.

As a criterion for choosing a valid $ T $, we require that it be much larger than the correlation time $ \tau $. To find an order of magnitude estimate for $ \tau $, we suppose the correlation function decays exponentially as $ R(n) \sim e^{-n / \tau} $. We expect the quantity $ \sum R(i) $ to be the same order of magnitude as $ \int e^{-x / \tau} dx = \tau $. And so from \eqref{approxvar} we estimate $ \text{Var}(X_1) \approx 2\sigma^2 \tau / T $. Thus the requirement that $ \tau \ll T $ leads to the condition
\begin{equation}
 \text{Var}(X_1) \ll 2\sigma^2 . 
\label{ineq}
\end{equation}

In practice we chose a sufficiently large $ T = 10,000$ and verified that the sample variance of $ X_1 $ satisfied \eqref{ineq} throughout.

We may also calculate the sample variances and covariances of the higher order coarse-grained variables $ X_2 $ and $ X_4 $. Using standard propagation of error methods, we may use these to derive formulas for the variances of $ \sigma^2 $ and the Binder cumulant $ U $,
\begin{eqnarray}
 \text{Var}(\sigma^2) & = & \text{Var}(X_2) + 4 \varphi^2 \text{Var}(X_1)\nonumber\\
 && -\: 4 \varphi \text{Cov}(X_1,X_2) \\
 \text{Var}(U) & = & (1-U)^2(\dfrac{\text{Var}(X_4)}{\langle X_4\rangle^2}
 + \dfrac{\text{Var}(X_2)}{\langle X_2\rangle^2} \nonumber\\
  && -\: 4 \dfrac{\text{Cov}(X_2,X_4)}{\langle X_2\rangle \langle X_4\rangle}) 
\label{varu}
\end{eqnarray}

\begin{acknowledgments}
The authors would like to thank Marie Lopez del Puerto and Gerry Ruch for their help in using the University of Saint Thomas Physics department's Beowulf cluster, which was instrumental in running these computer simulations.
\end{acknowledgments}

\bibliography{paperR}

\end{document}